# TRAJECTORY DESIGN FROM GTO TO LUNAR EQUATORIAL ORBIT FOR THE DARK AGES RADIO EXPLORER (DARE) SPACECRAFT

Anthony L. Genova,[*] Fan Yang Yang,[†] Andres Dono Perez,[‡] Ken F. Galal,[§] Nicolas T. Faber,[¶] Scott Mitchell,[#] Brett Landin,[**] Abhirup Datta[††], and Jack O. Burns[‡‡, §§]

The trajectory design for the Dark Ages Radio Explorer (DARE) mission concept involves launching the DARE spacecraft into a geosynchronous transfer orbit (GTO) as a secondary payload. From GTO, the spacecraft then transfers to a lunar orbit that is stable (i.e., no station-keeping maneuvers are required with minimum perilune altitude always above 40 km) and allows for more than 1,000 cumulative hours for science measurements in the radio-quiet region located on the lunar farside.

## INTRODUCTION

This paper presents a trajectory design whereby the Dark Ages Radio Explorer (DARE) spacecraft can be launched as a secondary payload into a geosynchronous transfer orbit (GTO) and through a series of maneuvers reach lunar orbit. The final lunar orbit does not require any maintenance maneuvers to keep perilune above a safe 40 km altitude, yet satisfies a science requirement to obtain 1,000 hours of cumulative observation time in a radio-quiet region located on the lunar farside to explore the origin and evolution of the first stars and galaxies of the universe.

The DARE mission is currently in the concept development phase and has been submitted in response to NASA's 2014 Small Astrophysics Explorer (SMEX) Program call for proposals. DARE team partners include NASA Ames Research Center (ARC), Ball Aerospace and Technologies Corp., the Jet Propulsion Laboratory (JPL), and the University of Colorado at Boulder.

Results of the DARE trajectory design and analysis are presented herein.

[*] Trajectory Designer, Mission Design Division, NASA Ames Research Center, Moffett Field, CA 94035
[†] Aerospace Systems Engineer and Orbit Analyst, Mission Design Division, Science and Technology Corporation, based at NASA Ames Research Center, Moffett Field, CA 94035
[‡] Aerospace Systems Engineer and Orbit Analyst, Mission Design Division, Universities Space Research Association (USRA), based at NASA Ames Research Center, Moffett Field, CA 94035
[§] Spacecraft Attitude, Orbit, and Systems Engineer, Programs and Projects Division, NASA Ames Research Center, Moffett Field, CA 94035
[¶] Project Manager and Aerospace Systems Engineer, Mission Design Division, Stinger Ghaffarian Technologies (SGT), based at NASA Ames Research Center, Moffett Field, CA 94035
[#] Aerospace Systems Engineer and Orbit Analyst, Ball Aerospace and Technologies Corp., Boulder, CO 80301
[**] Aerospace Systems and Spacecraft Design Engineer, Laboratory for Atmospheric & Space Physics, University of Colorado, Boulder, CO 80303
[††] Postdoctoral Research Associate, Center for Astrophysics and Space Astronomy, Univ. of Colorado, Boulder, CO 80303
[‡‡] Professor, Department of Astrophysical and Planetary Sciences, Univ. of Colorado Boulder, Boulder, CO 80309
[§§] Principal Investigator, NASA Lunar Science Institute, NASA Ames Research Center, Moffett Field, CA 94035

**RADIO-QUIET REGION DEFINITION**

Human-generated radio transmissions are the major source of radio frequency interference (RFI) in Earth orbit and above the near-side of the Moon (McKinley[1]). At the radio frequencies used for DARE (40-120 MHz), the Earth's ionosphere, with a plasma frequency of ~10 MHz, easily passes this RFI into space. Typical FM radio transmitters with a broadcast power of 100 kW produce emissions that are nearly $10^9$ times stronger than the cosmological signal that DARE will pursue.

To locate a region in space for DARE that provides shielding from such RFI, efforts were focused on the farside of the Moon (also considered by others such as Maccone[2,3]). Although the RFI is still ~$10^8$ times the brightness of the 21-cm Dark Ages signal at lunar distance, the farside of the Moon can shield a spacecraft from this RFI and allow its radiometer to operate in a low background environment to observe the weak emissions from the early Universe[4].

The prescribed radio-quiet region is very unique in the Solar System since the Moon acts as a shield not only from the emissions that come from Earth but also from the Sun. Radio-emissions from potential solar bursts and other solar activity can be a source of corruption to DARE's required science observations. It was also necessary to account for emissions from spacecraft in the GEO belt and diffraction effects produced by the spherical shape of the Moon during wave propagation. Figure 1 shows a representation of two separate shadow cones calculated by considering the emissions from both the Sun and the Earth (shadow cones shown in dark yellow and dark teal, respectively). The true radio quiet region is the volume intersected by these two cones, shown in (light) red on the lunar farside (Fig. 1). The DARE mission requires that more than 1,000 hours is accumulated in this radio-quiet region for science data collection over a two year period.

After calculating the geometric properties of the cone shapes, the Analytical Graphics Inc. (AGI) Systems Tool Kit (STK) was used to generate a computer based model that calculated the time that the spacecraft was in the radio-quiet region. The STK/Astrogator scenario utilized a High Precision Orbit Propagator (HPOP) and $7^{th}$ order Runge-Kutta numerical integrator, which included a LP150Q lunar gravity model. To speed up run time without compromising accuracy, a 70 degree and order lunar propagation model was used, while accounting for solar radiation pressure and third body perturbations from the Earth and Sun. A set of rotating coordinates was used to model the motion of the Earth and Sun quiet cones and establish the placement of their apex.

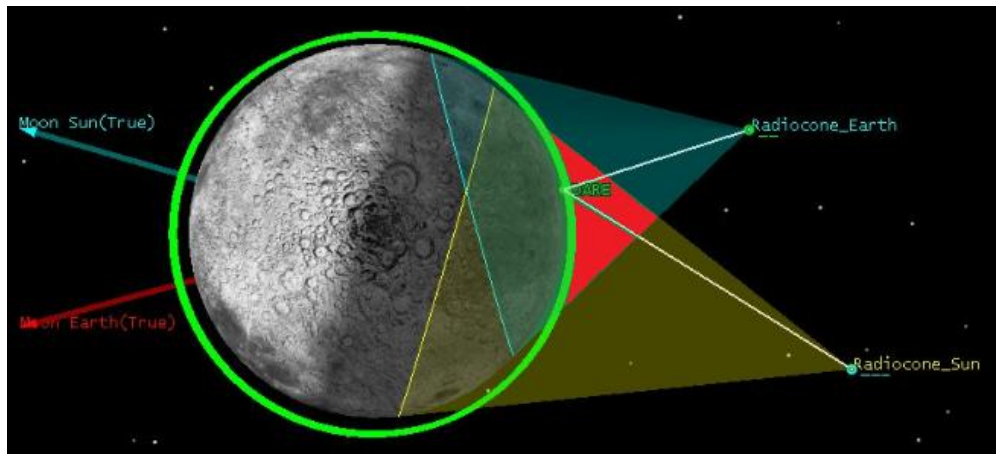

**Figure 1. DARE radio-quiet region for science observation seen as intersection (light red) of Earth (teal) and Sun (dark yellow) shadow cones on lunar farside, with view normal to lunar equator.**

## TRAJECTORY DESIGN & ANALYSIS

The trajectory analysis began by identifying acceptable ranges of lunar orbit altitude and inclination values that satisfied DARE's 1,000 hour observation requirement within two years. STK's Astrogator module and MatLab were the primary software tools used for the trajectory analysis. The Earth and Moon true-of-date (TOD) coordinate systems are used for all figures displaying relevant data such as inclination and right ascension of the ascending node (RAAN).

### Lunar Orbit Altitude and Inclination

In 2005, Ramanan & Adimurthy[5] studied the impact of inclination on the orbital lifetime of low altitude lunar orbits. A similar study was performed by the DARE team to determine the stability of all lunar orbit inclination values (0 to 180 degrees) in increments of one degree. The initial lunar orbits were assumed to be circular at 125 km altitude, and were propagated for two years. Four possible inclination ranges appeared to be stable (circled in Fig. 2) and were thus worthy of further exploration.

A lunar orbit inclination value was considered to be acceptable if it yielded a stable orbit (minimum perilune altitude above 40 km with no station-keeping maneuvers required) and more than 1,000 cumulative hours spent in the radio-quiet region within two years for all possible RAAN values of the initial lunar orbit. Four inclination ranges/values appeared to be stable and were thus studied in more detail: 0 to 1.4 degrees, 178.7 to 180 degrees, 25 to 28 degrees and 153 degrees (the latter is included in a narrow range of values although with a 38 km minimum perilune altitude). Inclination ranges significantly inclined to the lunar equator provide relatively little time in the radio-quiet region and were thus excluded for consideration.

Of the inclination ranges studied in detail, only one yielded more than 1,000 hours of science time (for all RAAN values) within two years and was thus chosen as baseline: **0 to 1.4 degrees**, (Figs. 3 and 4). Of note is the inclination ranges of 178.7 to 180 and 25 to 28 degrees yielded more than 1,000 hours of science time when the science duration was increased to 2.5 years.

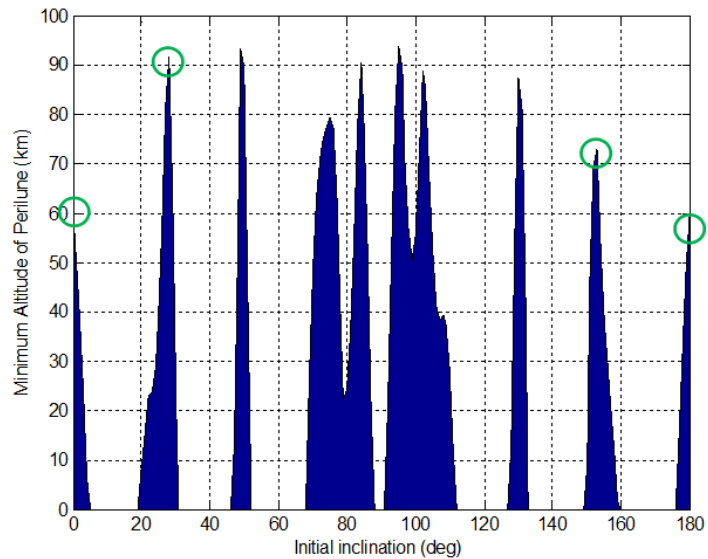

**Figure 2. Minimum Perilune Altitude vs. Lunar Orbit Inclination for an initially circular lunar orbit at 125 km altitude and 0 deg RAAN; 1 degree inclination increments and 2 year propagation.**

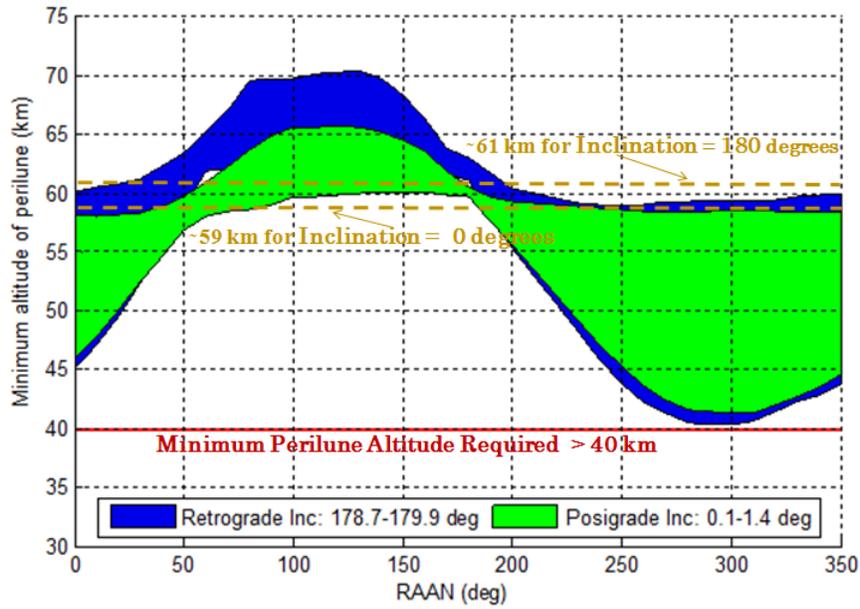

**Figure 3. Minimum perilune altitude vs. lunar orbit RAAN (10 degree increments) throughout 2 year science phase. Although represented as horizontal dashed lines, inclination values of 0 and 180 degrees have indeterminate RAAN values. Inclination increments of 0.1 degrees assumed.**

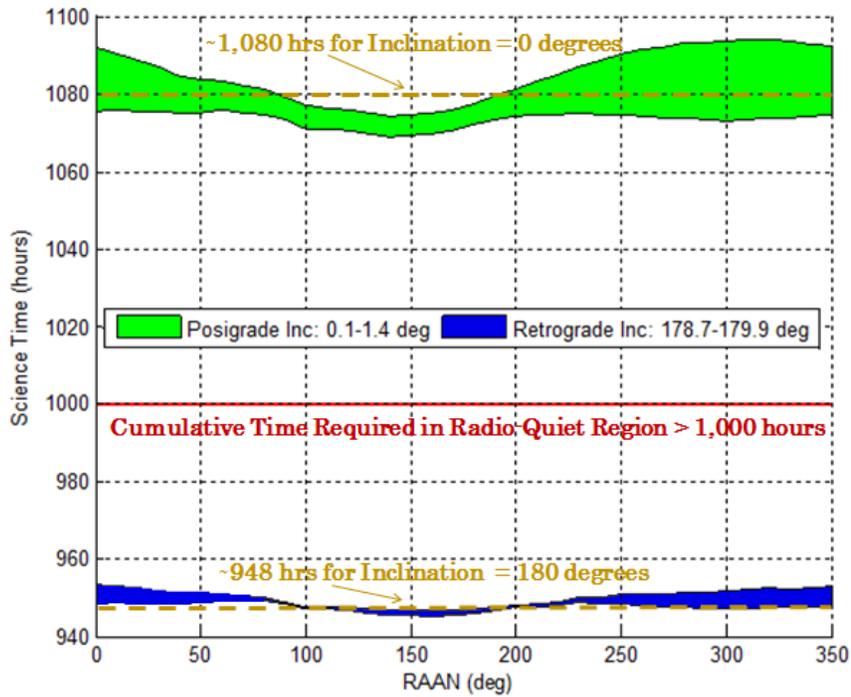

**Figure 4. Science time vs. lunar orbit RAAN (10 degree increments) after 2 years. Although represented as horizontal dashed lines, inclination values of 0 and 180 degrees have indeterminate RAAN values. Inclination increments of 0.1 degrees assumed.**

**Trajectory Design Method #1**

In order to minimize the ΔV cost to reach the Moon, the first trajectory design method considered was the use of a Sun-Earth weak stability boundary (WSB) transfer (Fig. 5). Since the nature of a WSB trajectory is affected by the orientation of the initial orbit (GTO) with respect to a fixed Sun-Earth line in a Sun-Earth rotating frame, assumptions were made regarding the argument of perigee and the local launch time of the insertion GTO. As pointed out by Qauntus et al.[6] typical Ariane V launches to GTO yield a geometry that is favorable for a WSB transfer back to the Moon: perigee is located in Earth's shadow with apogee in quadrant II (an apogee located in quadrant IV is also favorable for lunar-return but not assumed for this method)[7, 8]. This favorable geometry is evident from a sampling of the local solar time of perigee for historical GTO launches (relevant to the SMEX call for proposals) from 2011 to 2014 (Fig. 6), where a large cluster of perigee times is seen to occur near midnight (i.e., perigee in Earth's shadow). Thus for this trajectory design method, a GTO with a near-midnight perigee was used as in the initial condition to calculate Sun-Earth WSB transfers to the Moon for each day in a launch period from January 1 to December 31, 2020. The corresponding ΔV required to achieve a 125 km circular lunar orbit is shown in Figure 7 with ΔV across all launch dates totaling below 1,887 m/s.

It is of note that a value of 120 m/s was initially allocated for non-deterministic ΔV due to losses and trajectory correction maneuvers (TCMs). However, this value was later reduced to 90 m/s as the design matured and as such, all methods presented in this paper assume the 90 m/s value in their total ΔV requirement estimates and calculations, including the 1,887 m/s value above.

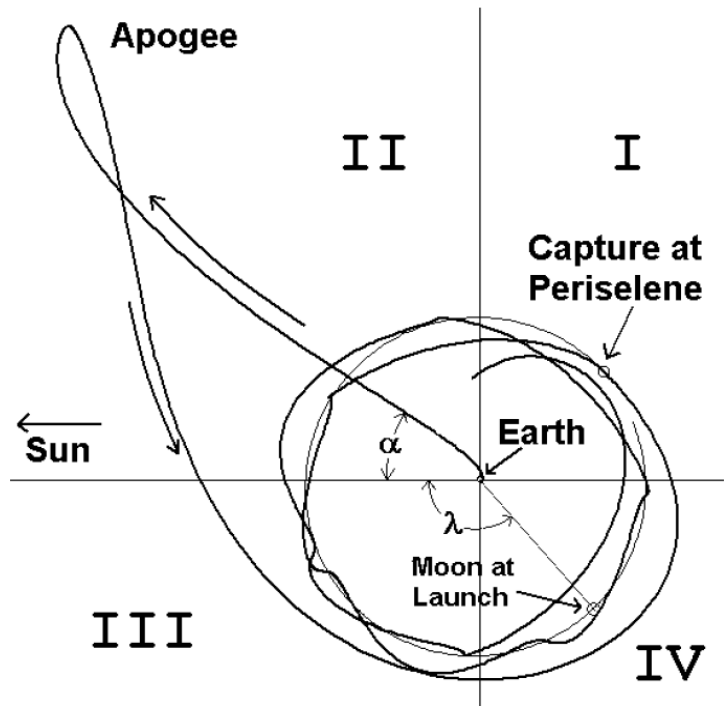

**Figure 5. Sun-Earth weak stability boundary (WSB) trajectory transfer shown with apogee in quadrant II, which is favorable for lunar-return (image from Ref. [7]).**

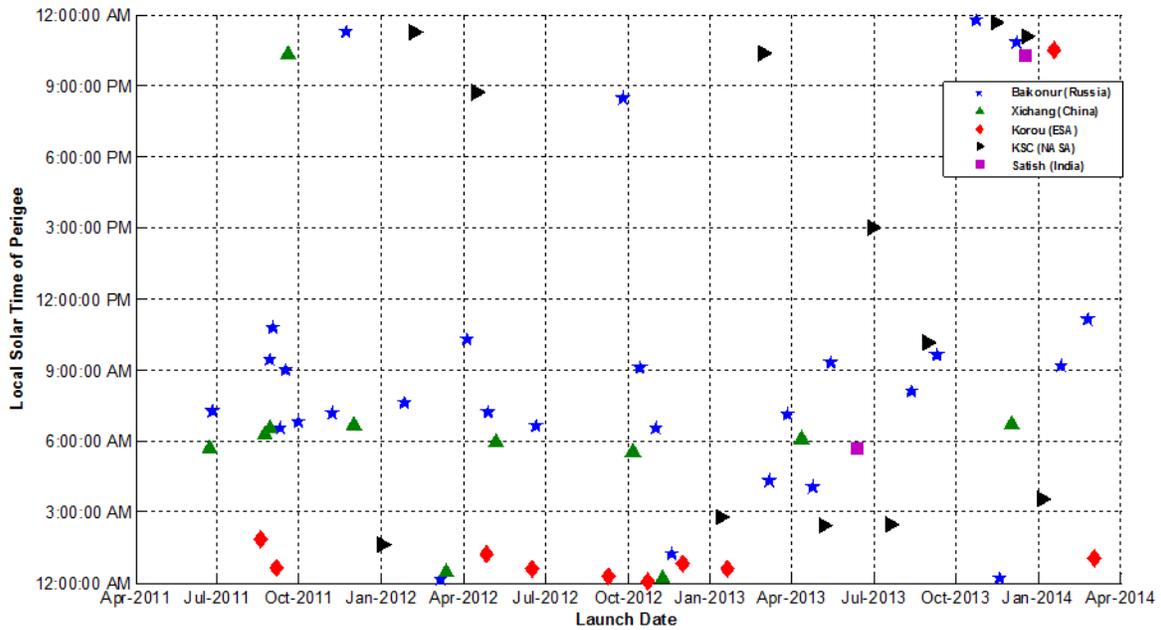

**Figure 6. Historical launches to GEO (via GTO) from April 2011 to April 2014. Data derived from launch times supplied by Analytics Graphics Inc. (AGI) of Exton, PA (manufacturer of STK).**

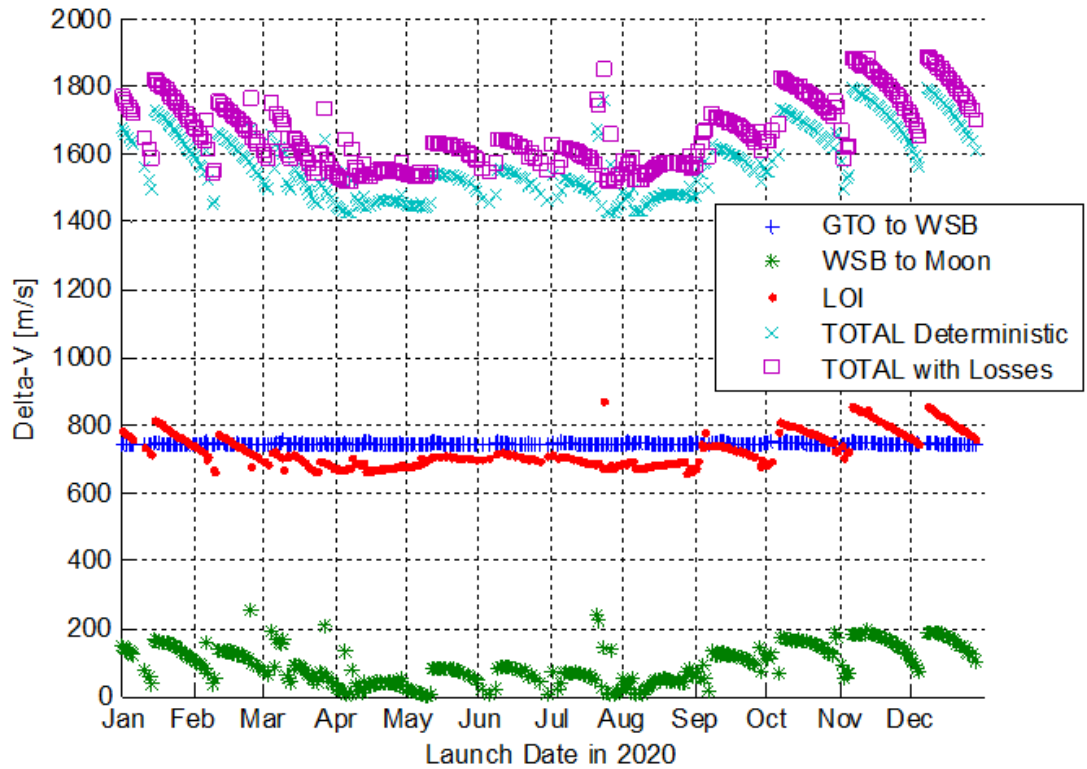

**Figure 7. ΔV requirements from GTO to the Moon using a Sun-Earth WSB transfer (method #1).**

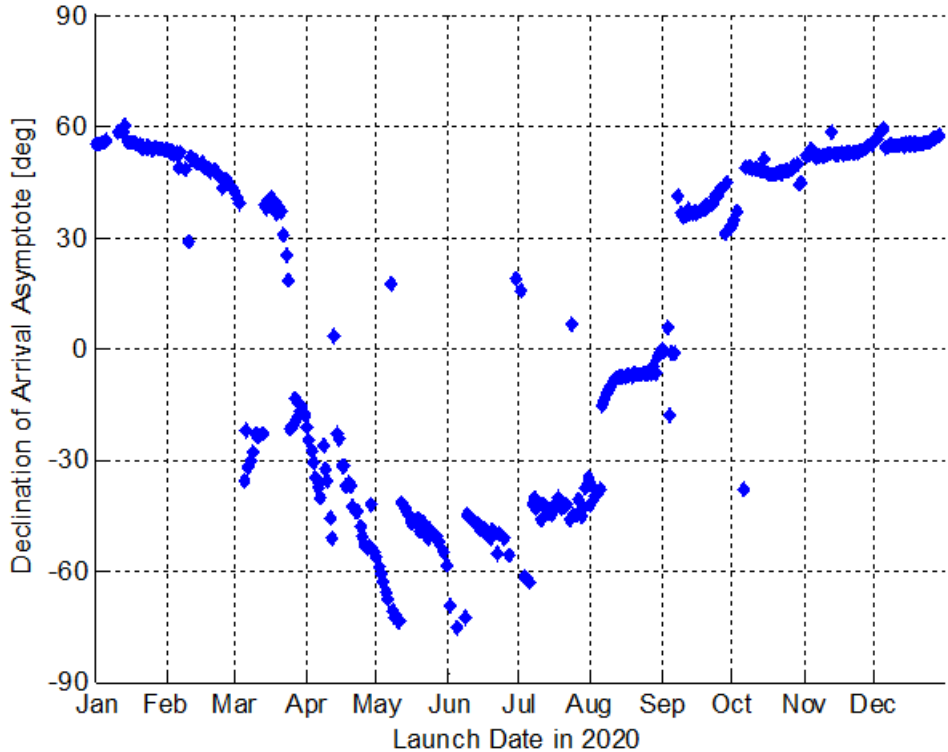

**Figure 8. Declination of the arrival asymptote (DAA) of the spacecraft's trajectory at lunar encounter directly after leaving the Sun-Earth WSB apogee.**

Unfortunately, many WSB transfer trajectories approach the Moon in orbit planes significantly inclined to the lunar equatorial plane. Figure 8 shows the declination of the arrival asymptote (DAA) at the Moon for the WSB trajectories calculated for DARE. Many of the cases have DAA values that result in lunar orbit inclinations significantly inclined to the lunar equatorial plane that do not satisfy the requirement for 1,000 hours of time in the radio-quiet region.

Therefore, an inclination change maneuver near apolune of the insertion orbit was utilized to target an acceptable value of inclination (nominally chosen as 26 degrees based on the 25 to 28 degree range that was assumed to be acceptable early in the design phase). Three intermediate lunar orbits with periods of 24, 48 hours and 72 hours were considered for the insertion orbit, with the 24 hour orbit selected primarily on the basis of relative orbit stability. Depending on the argument of perilune of the insertion orbit, the required inclination change ΔV could be as much as 800 m/s in this 24 hour period lunar orbit.

The resulting total ΔV, including 90 m/s for non-deterministic maneuvers and losses, exceeded the DARE spacecraft's ΔV capability of 2,065 m/s for many launch dates (Fig. 9). Therefore a more efficient way to change the lunar inclination was explored (see Trajectory Design Method #2).

Note that the current baseline plan is to launch DARE with its propellant tanks partially filled to minimize spacecraft mass and maximize the likelihood of a GTO rideshare opportunity. Thus the DARE spacecraft could fly this trajectory method by adding more propellant to its tanks.

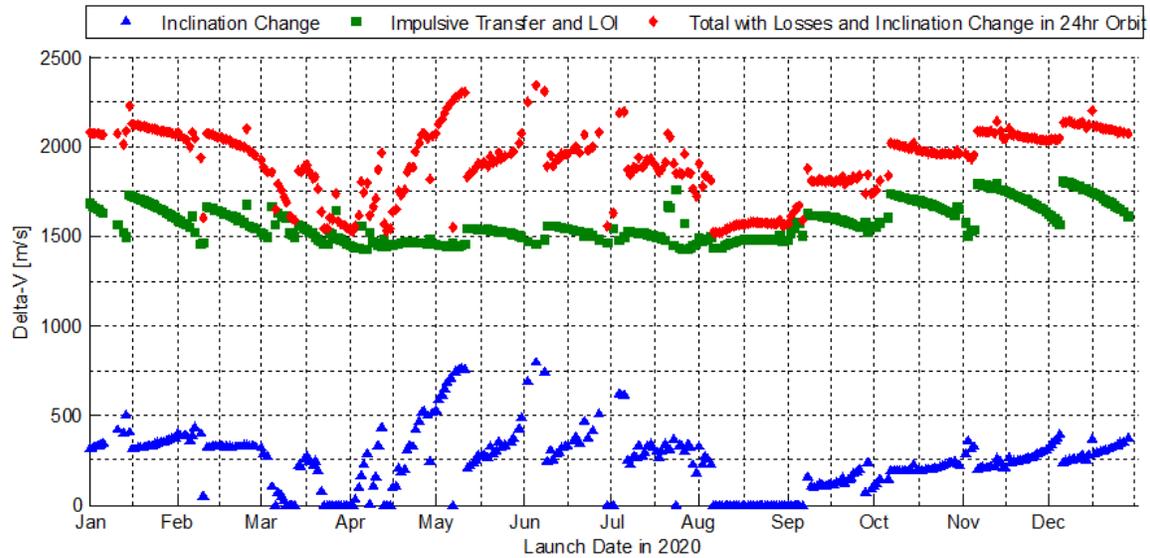

**Figure 9. Total ΔV requirement for trajectory method #1, including ΔV for plane-change maneuver in 24 hour period lunar orbit vs. launch date (calculated every day in calendar year 2020).**

**Trajectory Design Method #2**

A method of solving the lunar inclination problem encountered in Trajectory Design Method #1 involves using the initial lunar encounter targeted from the Sun-Earth WSB not as a chance to enter lunar orbit, but as a lunar flyby opportunity to change the spacecraft's orbital plane without the use of propellant.

Trajectories throughout the chosen analysis year of 2020 (from Fig. 7) were solved using this lunar flyby method to determine the solution that required the highest total ΔV. This solution is presented to demonstrate this trajectory design method: launch and separation from the primary payload in GTO occurs on December 6, 2020 (Fig. 10, A). After incrementally raising apogee (Fig. 10, B) to escape the Van Allen radiation belts (to limit radiation exposure and finite burn gravity losses) and phase with the Moon, the spacecraft is injected to an apogee altitude (Fig. 10, C) near the Sun-Earth WSB (1.4 million km from Earth). Upon reaching WSB apogee, a maneuver is performed (59 m/s of ΔV) to target a 2,900 km altitude flyby near the lunar south pole of the Moon on March 23, 2021. This modified approach contains a steep |DAA| of 72 degrees (compared to the 53 degree approach |DAA| in method #1) and thus benefits greatly from the targeted lunar flyby (Fig. 10, D), which rotates the orbital plane closer to the lunar orbit plane. The post-flyby Earth phasing orbit is in 2:1 resonance with the Moon (Fig. 10, E) and from the perigee (Fig. 10, F) of this orbit a maneuver is performed to target LOI on April 8, 2021 (Fig. 10, G). The subsequent lunar capture orbit is inclined 26 degrees to the equator (i.e., the preliminary baseline orbit inclination) and of 24 hour period. Two more maneuvers are performed at perilune (yielding a total LOI ΔV requirement of 712 m/s) to circularize the orbit and begin the science phase of the mission.

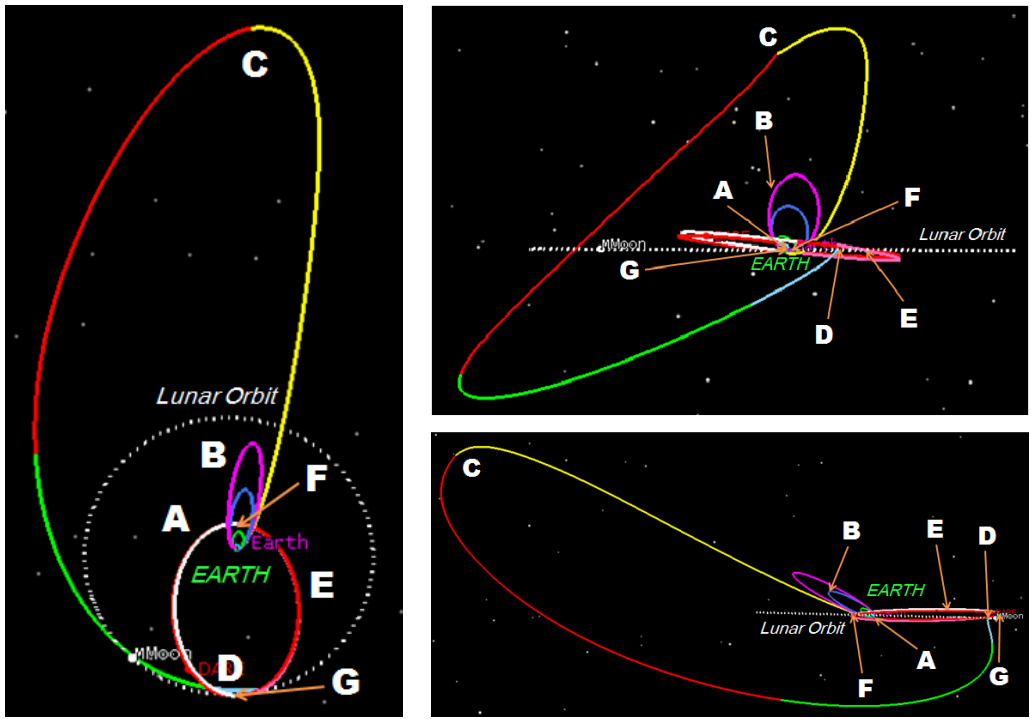

**Figure 10. Trajectory design from GTO to Lunar Orbit using a Sun-Earth WSB lunar transfer and lunar flyby. Trajectory shown in Earth inertial frames with view normal to (left) and edge-on (top-right and bottom-right) to the lunar orbit plane.**

The total deterministic ΔV for this case is 1,589 m/s. The additional ΔV of 90 m/s for non-deterministic maneuvers yields a maximum total ΔV requirement of 1,679 m/s.

Note that an inclination closer to the lunar equator could also have been targeted, with up to 75 m/s more ΔV needed to complete the plane change while in an elliptical lunar orbit (see Trajectory Design Method #3). To be conservative, the 75 m/s of maximum plane-change ΔV is added to the highest-ΔV (1,679 m/s) solution presented in Figure 10. The resulting maximum total ΔV requirement value 1,754 m/s will be used later for comparison purposes among methods.

Additionally, historical launch data was used to estimate the ΔV capability needed using this transfer approach to reach a low-inclination lunar orbit. For selected GTO launches between 2012 and 2014 (Table 1), the largest ΔV requirement for a hypothetical rideshare opportunity was about 1,670 m/s (for MUOS-2 simulated conditions with launch on an Atlas V). This result (Fig. 11) is similar to the previous estimate of 1,679 m/s (or 1,754 m/s) maximum required ΔV.

The possibility of eliminating launch time restrictions with trajectory design method #2 was analyzed as well. On a single launch date (February 19, 2014), trajectories were computed for all RAAN values of a GTO between 0 and 360 degrees, in one degree increments (Fig. 12). The spacecraft was assumed to wait in GTO until its apogee rotated into Sun-Earth WSB quadrant II or IV (due to the Earth's orbital motion around the Sun and J2 precession from Earth while in GTO phasing orbits). The maximum wait time is about eight months, which can be accomplished with little to no added ΔV. A possible way to decrease the wait time is to raise the apogee of the phasing orbits to nearly lunar distance or beyond (e.g., near the WSB) [9], but more analysis is needed to determine the number of phasing orbits (and their periods) that result in the fastest apogee rotation in the Sun-Earth rotating frame.

**Table 1. Selected historical data (2011 to 2014) for hypothetical DARE rideshare to GTO.**

| LAUNCH VEHICLE | NAME OF PAYLOAD | LAUNCH DATE | LAUNCH TIME |
|---|---|---|---|
| Atlas V (401) | TDRS 11 (TDRS K) | January 31, 2013 | 01:48 UTC |
| Atlas V (551) | MUOS 2 | July 19, 2013 | 13:00 UTC |
| Atlas V (531) | AEHF 3 (USA 246) | September 18, 2013 | 08:10 UTC |
| Atlas V (401) | TDRS 12 (TDRS L) | January 24, 2014 | 02:33 UTC |
| Delta-4M+(5,4) | WGS 4 (USA 233) | January 20, 2012 | 00:38 UTC |
| Delta-4M+(5,4) | WGS 5 (USA 243) | May 25, 2013 | 00:27 UTC |
| Delta-4M+(5,4) | WGS 6 (USA 244) | August 8, 2013 | 00:29 UTC |
| Falcon 9 v1.1 | SES 8 | December 3, 2013 | 22:48 UTC |
| Falcon 9 v1.1 | ThaiCom 6 | January 6, 2014 | 22:06 UTC |

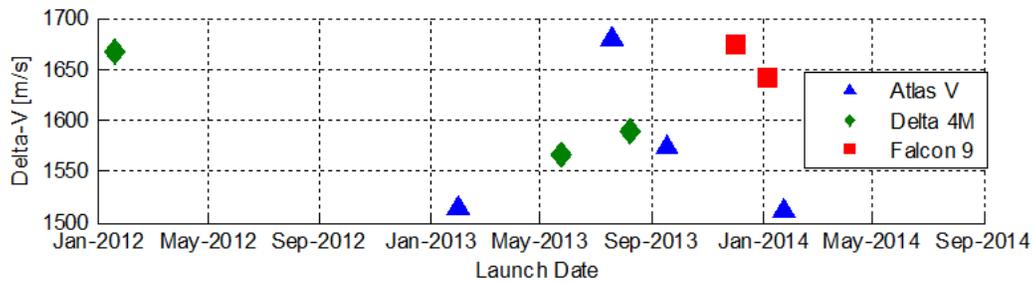

**Figure 11. Total ΔV requirements for DARE rideshare with selected launches to GTO (2011 - 2014).**

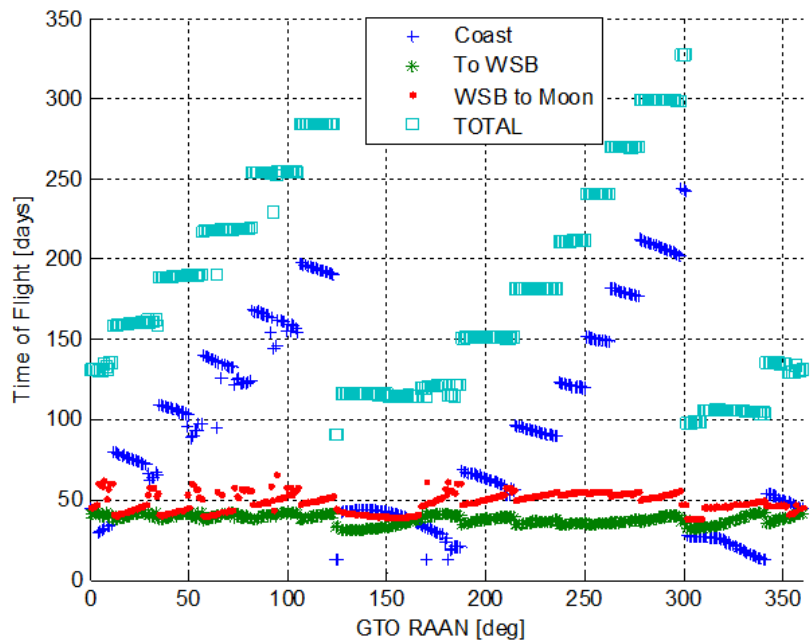

**Figure 12. Mission duration vs. RAAN (any time of day launch) for trajectory design method #2.**

**Trajectory Design Method #3**

As a final trajectory design option, Method #2 was modified by lowering the maximum apogee from near the Sun-Earth WSB to a super-lunar value of ~800,000 km (similar to that seen in Itoh[10], Ishii[11] and Tanabe et al[12]). This super-lunar apogee location is essentially shielded from Sun-Earth WSB gravitational effects, which simplifies the trajectory (and communications system) design at the expense of required ΔV as compared to trajectory design method #2.

To understand the effects of a GTO drop-off at any time of day, trajectories were first solved to the point of reaching lunar distance (projection onto lunar orbit plane) over a 360-degree range in RAAN in 30 degree increments (Fig. 13). After a further analysis of RAAN values of 10 and 211 degrees, it was seen that the spacecraft's orbit naturally crossed the lunar orbit plane en route to super-lunar apogee enabling a relatively short (i.e., less than one week) transfer to the Moon since no lunar flyby was needed. Therefore, DARE's analysis assumed a direct lunar transfer for solutions of RAAN equal to 0 and 210 degrees (Figs. 13 and 14). All other analyzed trajectories utilized a lunar flyby to place the spacecraft's Earth phasing orbit in the lunar orbit plane.

To demonstrate the lunar flyby solution (and the effect of a high |DAA| flyby), a trajectory that is highly inclined to the lunar orbit plane (as seen in Figs. 13 and 14) is chosen for detailed presentation. This trajectory corresponds to a RAAN value of 270 degrees and its full solution is presented in Figure 15. After separation from the primary payload in GTO (Fig. 15, A), the spacecraft incrementally raises apogee (via ΔV of 729 m/s) to beyond lunar distance, more than 800,000 km from Earth (Fig. 15, B). A maneuver (ΔV of 277 m/s) is performed near this super-lunar apogee to target a lunar flyby (Fig. 15, C) which places the trajectory in the Moon's orbit plane. However, the spacecraft's trajectory does not always cross the lunar equatorial plane at the time of the lunar flyby and thus a 0 degree lunar orbit inclination is not always directly achieved from the lone flyby. Instead, an inclination of up to 6.5 degrees is achieved since the lunar equatorial plane is inclined ~6.5 degrees to the lunar orbit plane. After the lunar flyby, an Earth phasing orbit (Fig. 15, D) is entered with apogee near lunar distance to allow for lunar intercept and orbit insertion. LOI (ΔV of 275 m/s) is performed at a 300 km perilune altitude (Fig. 15, E) to target a 24-hour period lunar orbit, two months after separation in GTO. Next, a second plane-change maneuver (ΔV of 8 m/s) is performed at the ascending orbital node to achieve a 0 degree inclination orbit (Fig. 15, F). Two maneuvers (sum ΔV of 556 m/s) are then performed at perilune (Fig. 15, G and H) to achieve the 125 km circular (equatorial) science orbit (Fig. 15, I).

After solving the trajectories corresponding to the remaining solutions (within the 360 degree RAAN range seen in Fig. 16), the total ΔV requirements were determined and are displayed in Figure 17. The variation in the total ΔV requirement among flyby solutions is primarily attributed to the varying ΔV cost of targeting the Moon from a super-lunar apogee maneuver. This maneuver is largest for solutions near 90 and 270 degrees of initial GTO RAAN, while it is negligible when the orbit naturally crosses the lunar orbit plane after leaving super-lunar apogee, the latter true for RAAN values of 22 and 163 degrees (Figs. 13 and 14). The highest total deterministic ΔV requirement among analyzed RAAN solutions resulted from the 270 degree RAAN case with a sum of 1,968 m/s, which includes ΔV for estimated finite burn losses and statistical components such as TCMs for launch injection errors (Fig. 17). This result is expected since there is a relatively large (277 m/s) ΔV cost required to target the Moon. The lowest total ΔV requirements (among lunar flyby trajectories) were yielded from solutions with RAAN values of 30 and 180 degrees. This result is also expected since these RAAN values are near the 22 and 163 degree values found to yield lunar transfer orbits with natural crossings of the lunar orbit plane. Given the DARE spacecraft's ΔV capability of 2,065 m/s, at least 98% ΔV margin (as calculated in Table 2) is yielded for any launch time of day.

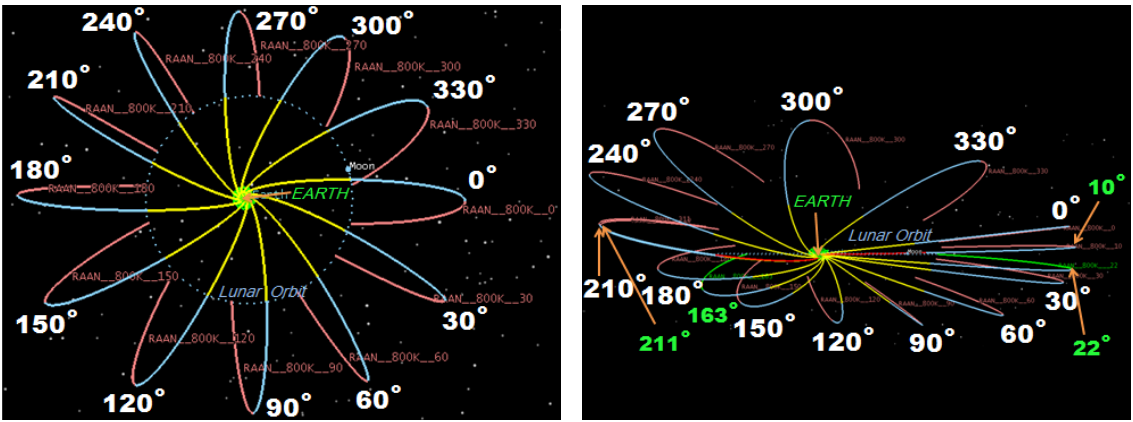

**Figure 13. Trajectory solutions from GTO to lunar distance (X direction in Earth-centered frame) for 360 degree range in RAAN in 30 degree increments, shown in Earth-centered, Earth inertial frames normal to (left) and edge-on (right) to the lunar orbit plane (lunar orbit dashed in light blue). Direct lunar transfers are feasible for RAAN values near 10 and 211 degrees (both labeled and seen in red crossing the lunar orbit plane, image on right). Low ΔV cost to set up lunar flyby for solutions near RAAN values of 22 and 163 degrees, which cross lunar orbit plane after apogee (both labeled and seen in green crossing the lunar orbit plane, image on right).**

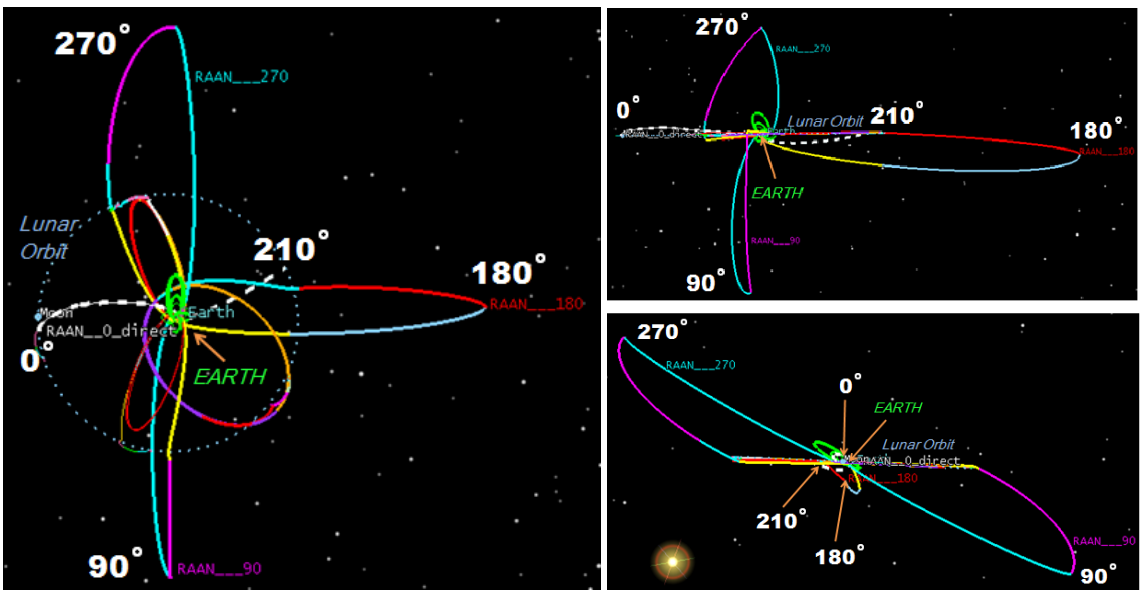

**Figure 14. Selected trajectory solutions for 0, 90, 180, 210, and 270 degrees of the initial GTO RAAN, shown in Earth-centered, Earth inertial frames normal to (left) and edge-on (top-right and bottom-right) to the lunar orbit plane (lunar orbit dashed in light blue). Direct lunar transfer solutions also shown as dotted white trajectories for relevant solutions corresponding to RAAN of 0 and 210 degrees.**

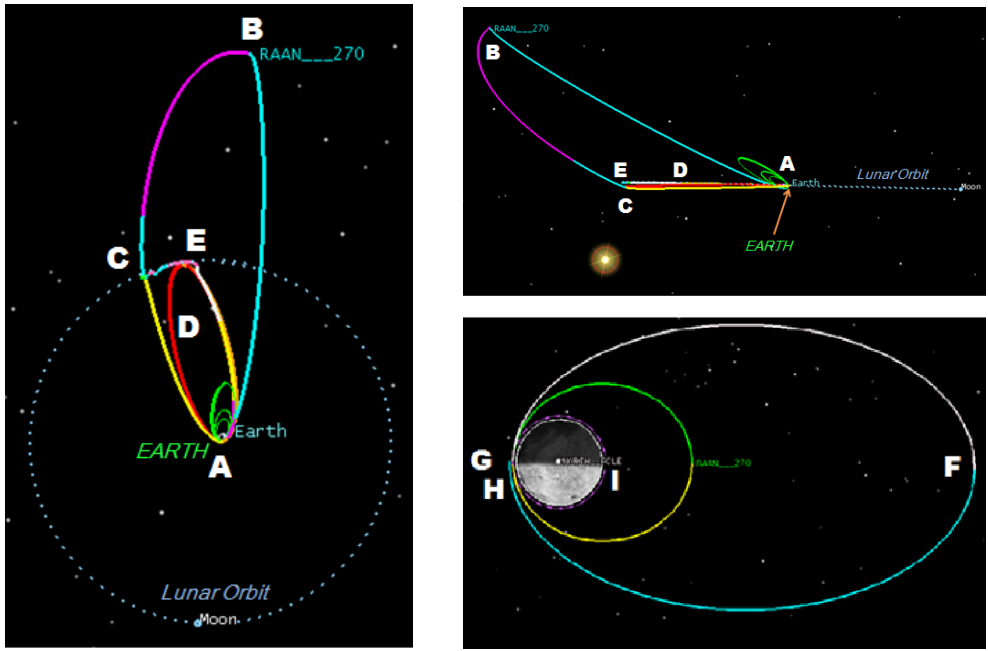

**Figure 15: Trajectory shown with initial GTO RAAN = 270 degrees, viewed in Earth-centered, Earth inertial frames normal to (left) and edge-on (top-right) to the lunar orbit plane (lunar orbit dashed in light blue). Trajectory is also shown in a Moon-centered, Moon-inertial frame, with a view normal to the lunar equatorial plane (bottom-right). Trajectory segments: A) GTO and phasing orbits, B) super-lunar apogee, C) Lunar Flyby Earth Phasing Orbit, E) Lunar orbit insertion (LOI) into a 24-hour period orbit used to change inclination to 0 degrees (from up to 6.5 degrees) via maneuver near the orbit node following after apolune (F). Acceptable lunar science orbit of 125 km circular and equatorial is achieved (I) after two perilune maneuvers (G & H).**

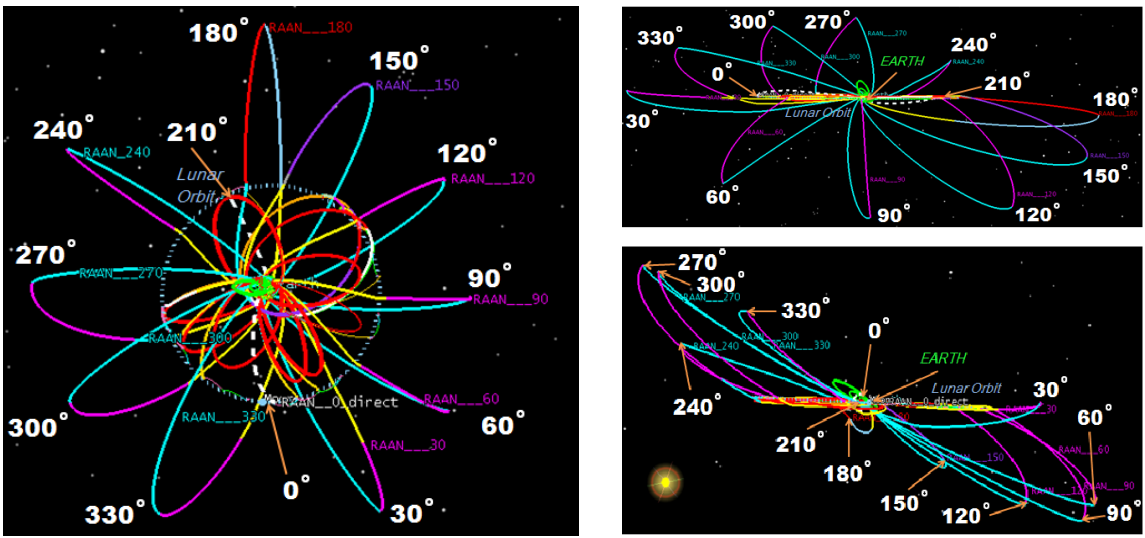

**Figure 16. Trajectory solutions for 360 degree range in RAAN in 30 degree increments, displayed Earth-centered, Earth inertial frames normal to (left) and edge-on (top-right and bottom-right) to the lunar orbit plane (shown dotted in light blue). Direct lunar transfer solutions are also displayed, dotted in white (0 and 210 degree RAAN solutions).**

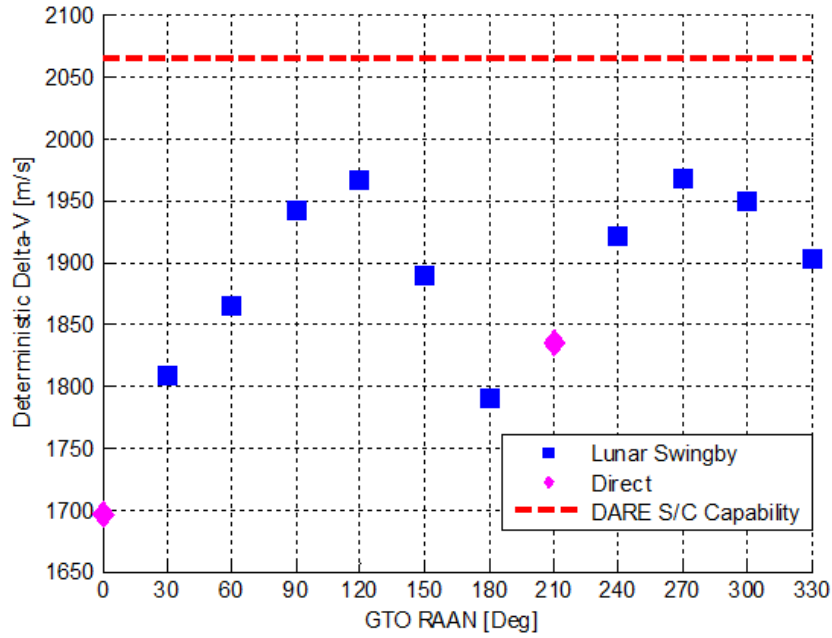

**Figure 17.** Survey of DARE deterministic ΔV for a GTO rideshare with RAAN values ranging from 0 to 360 degrees; 125 km circular baseline lunar orbit with 0 degree inclination assumed.

**Table 2. Summary of ΔV requirements for method #3 (270 degree RAAN solution shown, which contains the maximum total ΔV requirement among 360 degree range of analyzed RAAN values).**

| ΔV COMPONENT(s) | DETERMINISTIC ΔV (m/s) | STATISTICAL ΔV (m/s) |
|---|---|---|
| *Trajectory Changes*: maximum impulsive ΔV required for any time of launch, on any launch day | 1,927 | N/A |
| *GTO Perigee Maintenance* | 0 | 12 |
| *Launch Dispersions*: 3-sigma low injection error | 0 | 3 |
| *GTO Finite Burn Losses*: 3 maneuvers to raise apogee | 9 | 2 |
| *Trajectory Correction Maneuvers (TCMs)*, 3-sigma level | 0 | 29 |
| *LOI Finite Burn Losses*: 3 maneuvers to capture and lower apolune altitude at the Moon | 8 | 2 |
| *Lunar Orbit Phasing*: minor phasing maneuvers with Lunar Reconnaissance Orbiter (LRO) budgeted in the event of RF interference | 4 | 1 |
| *End of Mission Disposal*: impact with lunar surface | 20 | N/A |
| **Totals** | **1,968** | **49** |
| **MAXIMUM TOTAL ΔV Requirement** | colspan | 2,017 m/s |
| *DARE Spacecraft ΔV Capability* | colspan | 2,065 m/s |
| *ΔV Available for Statistical Components* | colspan | 97 m/s |
| *Percent ΔV Margin*: [2,065 – (1,968 + 49)] / 49 | colspan | **98%** |

## CONCLUSIONS

A stability analysis of lunar orbits was performed to determine candidate inclination ranges for DARE lunar science orbits. The initial design focused on the 25 to 28 degree stable inclination range which assumed a 2.5 year mission duration. Eventually, the more costly (in terms of ΔV) stable range of 0 to 1.4 degrees was chosen as the baseline since it yielded more than 1,000 hours of cumulative time inside the radio-quiet region within two years of science duration, as required.

Among trajectory design methods analyzed, method #3 contained the shortest average (i.e., mean) mission duration, about two months from GTO to LOI for lunar flyby solutions, while method #2 yielded the lowest maximum total ΔV requirement (~1,750 m/s). Method #3 was designed to accommodate a GTO launch at any time of day without increasing mission duration since its apogee is distanced enough from the Sun-Earth WSB (unlike in method #2). Although method #3's maximum total ΔV requirement (2,017 m/s) is relatively high, it is well within the DARE spacecraft's capability (yielding a minimum of 98% ΔV margin for any launch time of day, even with the spacecraft's tanks filled to only 78% of full capacity) and was thus selected as the baseline trajectory method for the DARE mission.

Although applied to a GTO launch, the presented trajectory designs are applicable to any spacecraft that needs to reach the Moon as a secondary payload dropped off in an Earth orbit with apogee below ~800,000 km. To utilize the presented methods, the final science orbit need not be equatorial (any lunar orbit inclination can be achieved), circular (elliptical orbits will result in ΔV savings) or an orbit for that matter since the spacecraft can choose to land or impact the lunar surface. A lunar lander may choose to perform its LOI at the time of the lunar flyby (methods #2 and #3) or after the nominally planned LOI (all three methods). Impact (or other) trajectories may also be targeted from the Sun-Earth WSB (methods #1 and #2) or super-lunar apogee (method #3) to take advantage of higher-declination approaches to the lunar orbit plane.